\documentclass[sigconf, nonacm]{acmart}

\AtBeginDocument{%
  }

\begin{document}

%%
%% The "title" command has an optional parameter,
%% allowing the author to define a "short title" to be used in page headers.
\title{Some Practice for Improving the Search Results of E-commerce }

%%
%% The "author" command and its associated commands are used to define
%% the authors and their affiliations.
%% Of note is the shared affiliation of the first two authors, and the
%% "authornote" and "authornotemark" commands
%% used to denote shared contribution to the research.
\author{Fanyou Wu}
\orcid{1234-5678-9012}
\affiliation{%
  \institution{Department of Forest and Natural Resource}
  \institution{Purdue University}
  \city{West Lafayette}
  \state{Indiana}
  \country{USA}
}
\email{wu1297@purdue.edu}

\author{Yang Liu}
\affiliation{
  \institution{School of Vehicle and Mobility \\ Tsinghua University}
  \city{Beijing}
  \country{China}
}
\email{liuy@chalmers.se}

\author{Rado Gazo}
\affiliation{
    \institution{Department of Forest and Natural Resource}
  \institution{Purdue University}
  \city{West Lafayette}
  \state{Indiana}
  \country{USA}
}

\author{Benes Bedrich}
\affiliation{%
  \institution{Department of Computer Science}
  \institution{Purdue University}
  \city{West Lafayette}
  \state{Indiana}
  \country{USA}
}

\author{Xiaobo Qu}
\affiliation{%
  \institution{School of Vehicle and Mobility Tsinghua University}
  \city{Beijing}
  \country{China}
}

%%
%% By default, the full list of authors will be used in the page
%% headers. Often, this list is too long, and will overlap
%% other information printed in the page headers. This command allows
%% the author to define a more concise list
%% of authors' names for this purpose.
\renewcommand{\shortauthors}{Wu et al.}

%%
%% The abstract is a short summary of the work to be presented in the
%% article.
\begin{abstract}
In the Amazon KDD Cup 2022, we aim to apply natural language processing methods to improve the quality of search results that can significantly enhance user experience and engagement with search engines for e-commerce. We discuss our practical solution for this competition, ranking 6th in task one, 2nd in task two, and 2nd in task 3. The code is available at \url{https://github.com/wufanyou/KDD-Cup-2022-Amazon}.
\end{abstract}




\begin{teaserfigure}
  \includegraphics[width=\textwidth]{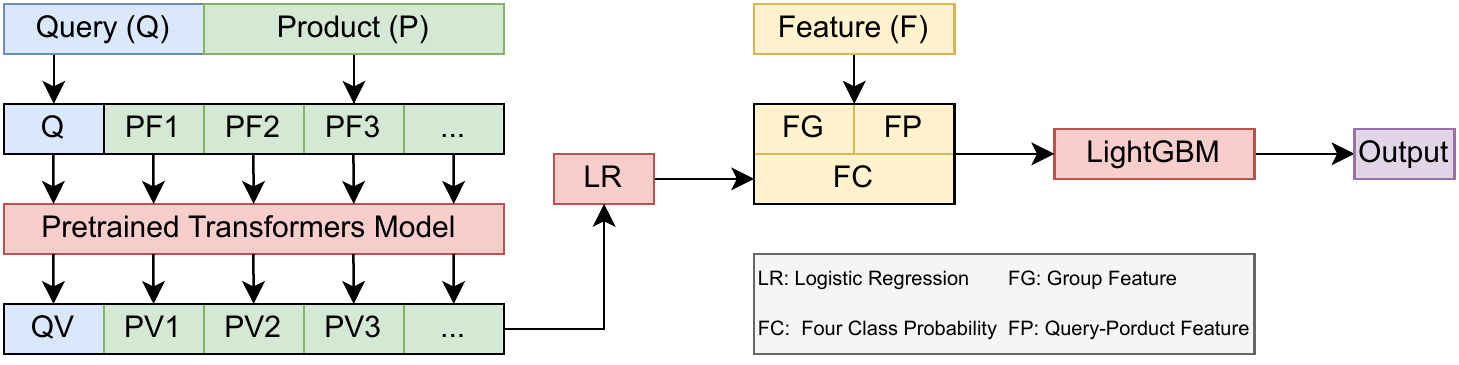}
  \caption{\label{fig:overall}Overall schema of our proposed solution for Amazon KDD CUP 2022 for all three tasks.}
  \label{fig:teaser}
\end{teaserfigure}

\maketitle

\section{Problem Description}
The organizer provides a dataset called \textit{the Shopping Queries Dataset}~\cite{reddy2022shopping}. It is a large-scale, manually annotated dataset composed of challenging customer queries. The data is multilingual and includes English, Japanese, and Spanish queries. It comprises query-result pairs with annotated four classes of relevance (ESCI labels):
\begin{itemize}
    \item \textbf{Exact (E):} the item is relevant for the query and satisfies all the query specifications; 
    \item \textbf{Substitute (S):} the item is somewhat relevant: it fails to fulfill some aspects of the query, but the item can be used as a functional substitute;
    \item \textbf{Complement (C):} the item does not fulfill the query but could be used in combination with an exact item; 
    \item \textbf{Irrelevant (I):} the item is irrelevant, or it fails to fulfill a central aspect of the query.
\end{itemize}

The primary objective of this competition is to build new ranking strategies and, simultaneously, identify interesting categories of results (i.e., substitutes) that can be used to improve the customer experience when searching for products. The three different tasks for this KDD Cup competition using our Shopping Queries Dataset are:
\begin{enumerate}
    \item[T1.] Query-Product Ranking
    \item[T2.] Multiclass Product Classification
    \item[T3.] Product Substitute Identification
\end{enumerate}

Task one (T1) aims at ranking the relevance of \textbf{a subset of the ESCI dataset} by using \textbf{Normalized Discounted Cumulative Gain (nDCG)} score to measure the performance. The organizer designed a \textbf{customized Discounted Cumulative Gain (DCG) of 1.0, 0.1, 0.01 and 0.0,} for Exact, Substitute, Complement and Irrelevant respectively. 

Task two (T2) aims to classify each product as being an Exact, Substitute, Complement, or Irrelevant match for the query. \textbf{The Micro-F1 (equivalent to accuracy here)} will be used to evaluate the methods. Task three (T3) is a binary classification problem that tries to distinguish whether a query product pair is a Substitute or not and uses the Micro-F1 score as well to measure the performance.

T1 uses a subset of ECSI dataset, while T2 and T3 use the same dataset. It is natural to treat this competition as two different problems. In the rest of the paper, we will use short-term \textbf{T2T3} to represent tasks two and three. This inner difference among tasks is also reflected on the final leaderboard that the final ranking of T1 and T2T3 are not correlated well. The setup also involves a potential \textbf{data leakage}, which we will discuss further in section~\ref{sec:EDA}.

\section{Explore Data Analysis}\label{sec:EDA}
ECSI dataset contains three tables, which are 
\verb|product_catalogue|, \verb|train| and \verb|test|. The \verb|test| is also decomposed into \verb|public| and \verb|private| where the latter one is unseen to us. In this section, we make our special observation of \verb|product_catalogue| and \verb|train| that play an important role in the final leaderboard.

\begin{figure}[hbt]
  \centering
  \includegraphics[width=\linewidth]{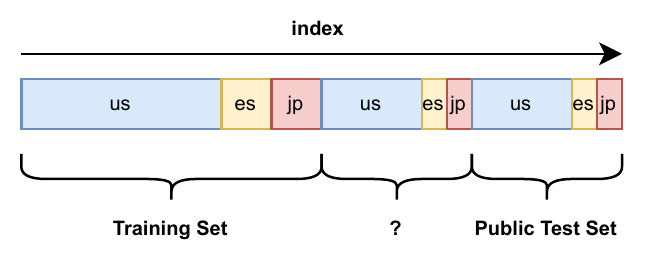}
  \caption{\label{fig:product}The order of product entries in T2T3.}
\end{figure}

T1 and T2T3 use different \verb|product_catalogue| tables. Figure~\ref{fig:product} shows the order of product entries in T2T3. T1 remains a similar pattern unless it is started with es entries (es $\rightarrow$ us $\rightarrow$ jp). There is a part of products that have not been used in the both training set and the public test set, and we conjecture those entries are unique in the private test set.

\begin{figure}[hbt]
  \centering
  \includegraphics[width=\linewidth]{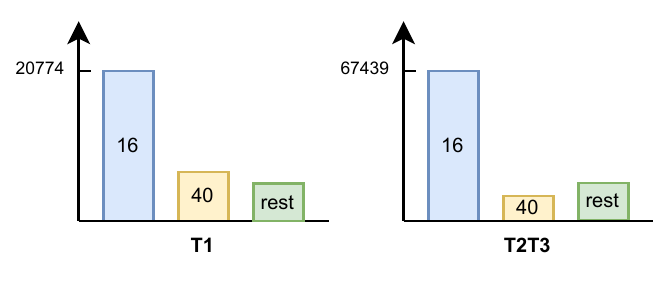}
  \caption{\label{fig:train}The histogram of products grouped by queries in the training set. 16 and 40 are the typical number of products for each query.}
\end{figure}

Let us have another investigation about \verb|train|. Figure~\ref{fig:train} shows the histogram of products grouped by queries in T1 and T2T3. Generally, most queries will sample 16 and 40 products for training and test sets. And this distribution is slightly different between T1 and T2T3, while we know that there are fewer Exact labels in T1. So associated with this prior knowledge, using this product number as a feature to calibrate the prediction will make some improvement in the leaderboard.

\begin{figure}[hbt]
  \centering
  \includegraphics[width=\linewidth]{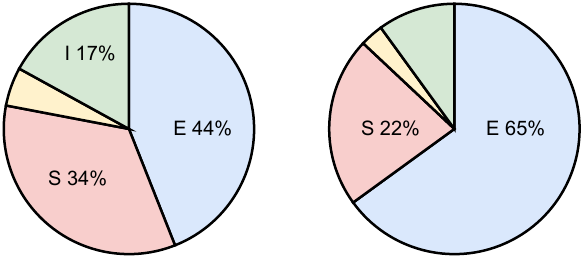}
  \caption{\label{fig:dist}The proportion of ESCI labels in T1 and T2T3 training set.}
\end{figure}

Another important piece of information in the training set is that the proportion of ESCI labels in T1 and T2T3 are different (Figure~\ref{fig:dist}). This difference creates a well-known data leakage for most participants in T2T3 that distinguishing whether a query-product pair is in T1 will improve results. However, it isn't easy to use this information directly in the private test set. 

Besides the above-described deep dive, some other patterns will help to improve the scores. Here is the summary of that information: 
\begin{itemize}
    \item[E1.] The order of the product catalog is not randomized (training set $\rightarrow$ private test set $\rightarrow$ public test set).
    \item[E2.] Most products are used once.
    \item[E3.] The ESCI label proportion is different between T1 and T2T3.
    \item[E4.] Most queries have 16 or 40 product numbers and the label distribution of those queries are slightly different.
    \item[E5.] The product id is called ASIN and will be identical to ISBN (starts with digits) if the product has ISBN.
    \item[E6.] Most query products group has fewer unique brand numbers than product numbers and the product with the most frequent brand tends to be labeled as Exact.
    \item[E7.] At least one product in a query-products group will be labeled as Exact, and the label of the query-product pair is affected by other labels in this group as well.
\end{itemize}

Combining those explorations, we could significantly improve scores for all three tasks. A detailed feature engineering will be introduced in Section~\ref{sec:feat}.

\section{Proposed Solution}
\begin{figure*}[hbt]
  \centering
  \includegraphics[width=\linewidth]{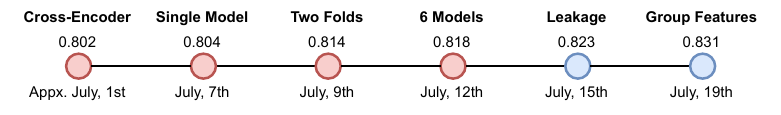}
  \caption{\label{fig:milestone} Our milestone of public leaderboard score for T2. We use red points and blue points to represent the improvement from cross-encoders models or the gain from feature engineering, respectively.}
\end{figure*}

Figure~\ref{fig:overall} shows the general schema of our proposed solution for Amazon KDD CUP 2022 for all three tasks. As we planned to attend to all three tasks, for efficiency, we have to train the cross-encoders once and use them for all three tasks. This strategy makes this two-stage solution the only choice. So we trained all cross-encoders with all data from T1 and T2T3 in two folds and then combined the four class probabilities with other essential features, using lightGBM to fuse and calibrate the prediction and adapt results to different tasks. Figure~\ref{fig:milestone} shows the milestone of the public leaderboard score for T2. In the following, we will discuss those milestones in detail.

\subsection{Cross-Encoder Architecture}
In the first stage, we applied the classical cross encoder~\cite{reimers2019sentence} architecture with only minor modifications. As the product context has multiple fields (title, brand, and so on), we use neither the CLS token nor mean (max) pooling to get the latent vector of the query-product pair. Instead, we concatenate the hidden states of a predefined token (query, title, brand color, etc.). This small modification yields about a 0.002 increase in the T2 public leaderboard (0.802 $\rightarrow$ 0.804).

In the final solution, we ensembled three cross-encoders for each language that differ from pre-trained models, the training data, or the input fields. For Engish entries, we used DeBERTaV3~\cite{he2021debertav3}, BigBird~\cite{zaheer2020big} and  COCO-LM~\cite{meng2021coco}. While for Japanese and Spanish ones, we used a multi-language version of DeBERTaV3. All those pre-trained data could be found at Huggerface. Table~\ref{tab:pre} shows the average accuracy for each model for T2. By ensemble of all models,  the score for the public leaderboard for task two is 0.818.
\begin{table}[hbt]
  \caption{Average accuracy for each model for T2. Here we used 2-fold cross-validation. Note for code submission, we used the different seeds to split data, so the accuracy here is not very comparable.}
  \label{tab:pre}
  \begin{tabular}{llr}
    \toprule
    \textbf{Locale} & \textbf{Pretrained Model} & \textbf{Accuracy} \\
    \midrule
    us & bigbird-roberta-base  & 0.7587\\
       & cocolm-base  & 0.7588\\
       & debertav3-base & 0.7585\\
    \midrule
    es  & mdeberta-v3-base & 0.7347 \\
    \midrule
    jp  & mdeberta-v3-base & 0.7006 \\
    \bottomrule
  \end{tabular}
\end{table}

\subsection{Feature Engineering}\label{sec:feat}
Once stacking a lot of models has tiny improvements for both local and online tests, we start to do some feature engineering based on our exploration in section~\ref{sec:EDA}.

\subsubsection{Leakage Features}
Combining E1, E2, and E3 together, we designed a feature that measures the percentage of \verb|product_id| in Task 1 product list grouped by \verb|query_id|. This feature gives us a 0.005 improvement in the T2 public leaderboard and remains effective for the private test set, and that's why we are extremely closed to first place in T2 (0.0001 difference).

\subsubsection{Query product number features} Based on E4, we use the product number for each query as a feature and obtain an approximate 0.002 improvement in the T2 public leaderboard. 

\subsubsection{Product ID features} Based on E5, we designed features that measure whether the \verb|product_id| is ISBN or not and whether the query-products group has an ISBN product or not. This feature gives us an approximate 0.001 improvement in the T2 public leaderboard.

\subsubsection{Brand Features} Based on E6, we designed features that measure the unique number of brands in a query-products group and whether the brand of the product is the most frequent one in the group. This feature gives us an approximate 0.001 improvement in the T2 public leaderboard.

\subsubsection{Group Features} Based on E7, we designed several stats (min, medium, and max) of the cross encoder output probability grouped by \verb|query_id|. This feature gives us a 0.008 improvement in the T2 public leaderboard.

\subsection{LightGBM model}
\subsubsection{T1} As the ECSI label distribution is different in T1 and T2T3, for T1, we train the lightGBM model with T1 data only to simulate this distribution and calculate the expected gain for each query-product pair as:
\begin{equation}
     s = p_e + 0.1 \cdot p_s + 0.01 \cdot p_c, 
\end{equation} 
where $p_e$, $p_s$ and $p_c$ are the probability output of lightGBM for label Exact, Substitute and Irrelevant respectively. Then we sort the query-product list by this gain. This method is slightly better than using LambdaRank~\cite{burges2006learning} with the same label gain (0,0.01,0.1,1) in LightGBM.

\subsubsection{T2T3} We use full data from T1 and T2T3, and use lightGBM to train either a four-class classifier (T2) or a binary classifier (T3). 

\subsubsection{Model ensemble} For T1, T2, and T3, we average the gain or the prediction from 6 models (3 models x 2 folds) for each language to make a final decision.

\section{Inference Acceleration}
During the code submission round, the organizer proposed a 120 minutes time limit for all three tasks. This time limit requests us to provide a more efficient solution.

\subsection{Knowledge Distillation}
We use knowledge distillation~\cite{hinton2015distilling} to improve the model's performance. This knowledge distillation is applied to English entries only. We used all data and trained large versions of DeBERTaV3, BigBird, and COCO-LM. Then we used a linear combination loss of cross-entropy (loss between prediction and ground truth) and mean-square-error (logit difference between student model and teacher model).

\subsection{Other Inference Acceleration Strategies}

Here are we listed some other inference acceleration strategies:

\begin{itemize}
    \item[A1.] Pre-process product token and save it as an HDF5 file.
    \item[A2.] Transfer all models to ONNX with FP16 precision.
    \item[A3.] Pre-sort the product id to reduce the side impact of batch zero padding.
    \item[A4.] Use a relatively small mini-batch size ($=4$) when inference.
\end{itemize}

\bibliographystyle{ACM-Reference-Format}
\bibliography{acmart.bib}
\end{document}